# SEARCH FOR CORRELATIONS BETWEEN MORPHOLOGICAL CHARACTERISTICS AND THE CRYSTALLITE SIZES IN THIN ZINC OXIDE FILMS

A. Kh. Argynova, A. A. Loctionov, K. A. Mit', D. M. Mukhamedshina

Institute of Physics and Technology, Almaty, Kazakhstan

**Abstract**

Size-property relations in plasma-modified ZnO thin films have been investigated as a function of plasma treatment duration. The correlations between crystallite sizes and the morphological characteristics of films have been extracted on the basis of the frontier computational analysis of the scanning probe microscope (SPM) data matrices. The nanocluster structure of oxide films have been studied in detail with high accuracy. The strong plasma-induced changes in crystallite sizes have been interpreted as a size-structure phase transition. Direct measurements of X-ray diffraction and optical transmittance spectra have confirmed the results obtained with computational approaches. The discovery size-morphology correlations in thin oxide films might open new avenues ultimately leading towards deeper insight into unsolved problems of evaluation of optimal technological conditions for thin oxide film designing.

**Introduction**

Metal oxide nanostructures have attracted great interest due the large variety of physical properties they present.  Control of size, surface and assembly properties of nanoscale materials are crucial steps towards their study in fundamental research and  implementation in high technologies [1]. The importance  of transparent conducting oxides (TCO) in the field  of future production of electric energy by sun light conversation has become unquestionable over the past time, as using TCO for front contact or intermediate reflector are  one of the technological key points of solar cell designing [2]. While a considerable amount of research has been devoted to optimizing the individual functional properties which determine the performance as TCOs, i.e. resistivity and optical transparency [3], considerable less work has been done on systematically studying the structure/property relation in nanostructured oxides. For these kinds of applications, it is a matter of primary importance to understand the relation between crystallite sizes, structure and properties of thin film oxides.

In this work the crystallite size distributions were studied in relation to the exposure time of hydrogen plasma on zinc oxide films, as plasma treatment is an effective tool of modification of synthesized film properties [4-6].  The analysis results of metal oxide thin film synthesis requires research of electrophysical, optical, morphological and structural films' features, i.e. requires using of the different measuring platform collection and consequently substantial time and financial resources. The estimation of size and morphology correlation is a way to create a nanolaboratory on the single measuring platform [7] by means of mathematical and computing tools, which would enable to obtain a set of morphological and structural films' features. The crystallite sizes in thin films (in the limiting

case – two-dimensional structures), from general point of view, should be recognized in morphological features of the surface. These experimental researches of the size-morphology correlations in plasma-modified ZnO thin films can be used for designing thin oxide films with optimal properties. Detection and recognition of these very weak correlations requires precise measurement tools and effective analysis methods. In this work a high precision set of digital XYZ coordinate matrices of films was obtained on JEOL-5200 scanning probe microscope (SPM). The correlation estimates were obtained on the basis of effective methods of spectral and clustering analyses. All computations were performed on the groundwork in the field of scanning probe microscopy [8-10], neural network computing [11-14], wavelet analysis [15-17] in the environment of MATLAB software platform [17].

1. Basic Data

Zinc oxide films with good chemical homogeneity were obtained by deposition from zinc acetate dehydrate solution ($Zn(CH_3COO)_2 \cdot 2H_2O$) dissolved in isopropyl alcohol ($C_3H_7OH$) with addition of the ammonium hydroxide ($NH_4OH$). The films were prepared by spin-coating methods on a glass substrate. To study the effect of plasma treatment time with the film ZnO substrate was cut into four pieces. The first part was left unchanged as control sample and the rest three were treated in hydrogen glow-discharge plasma during 5, 10 and 15 min respectively. Hydrogen plasma are generated at the pressure of 6.5 Pa with capacitive coupled r.f power (27.12 MHz) of 12.5 W. Temperature of processing of films did not exceed 100°C.

The measurement matrices of all samples in SPM had XY dimensions 512 x 512.

2. Extraction of Crystallite Size Distributions from Morphological Features of Thin Films

Rearrangements in metal oxide thin films arising at plasma treatments correspond to the complex processes. Their characteristic properties can be reflected in the morphology perturbations. These responses are too weak in order to change the morphology pattern as a whole. However, they can modulate the corresponding frequency or scale range on height series of studied surface. We intend to estimate the detectability of target correlation phenomena using the coordinate-scale pattern features. The specifications of the fine structure analysis have to be very carefully fitted. The best study of the non-stationary data, where the coordinate of event occurrence is important to the description of the phenomenon, can be performed with computation of the local energy density of signal in the wavelet transform. The significance of this approach is obvious. This quantity reveals how much each scale contributes at any given point in coordinate, so evolution in the coordinate is observable. It can be viewed as a microscope for looking into the coordinate-scale signal characteristics.

The algorithms of the continuous wavelet transform [15, 16] perform decomposition of data using a set of well localized wavelets. The contribution of each wave *W(a,b)* depends on the wave scale *a* and its coordinate *b*. The result is demonstrated as a wavelet spectrum - "the surface in the 3D space" with amplitude *W(a,b)*, scale on the vertical axe *a*, and coordinate on the horizontal axe *b*. This approach to the crystallite size estimation procedure is schematically presented in the left part of FIG. 1.

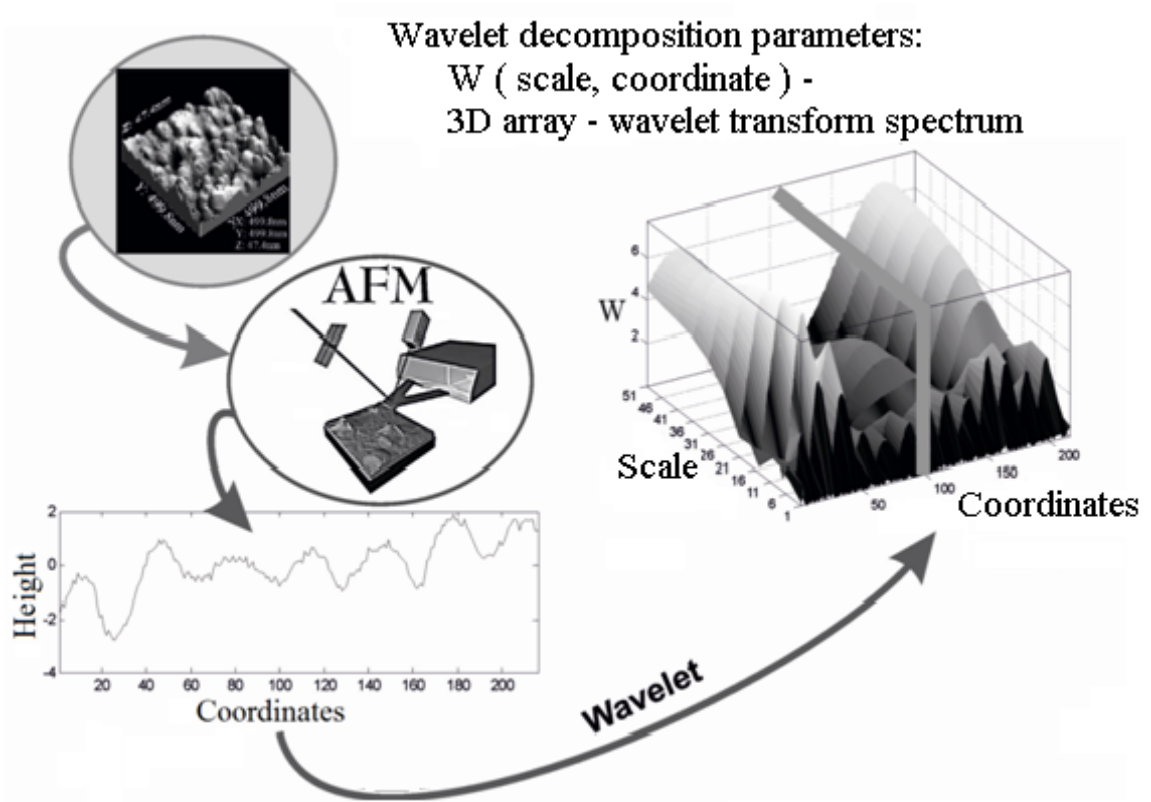

FIG. 1. Wavelet Analysis Scheme and Results

The estimated wavelet spectrum for the film after 10 min plasma treatment is presented in the right part of FIG.1. The scale estimations are based on the analysis of non-stationary height series f(z) on selected linear interval of the studied film surface. We have focused on a range of scales from a few to tens nanometers. The accuracy of crystallite size recognition in this approach corresponds to the results of the SPM with a tenfold sharper cantilever tip [19]. Computing is performed on the basis of wavelet analysis software tools in the MATLAB environment [17].

For this variant of plasma treatment the well-isolated bimodal scale pattern has been obtained in which the distinct narrow peak centered at approximately five nanometers is close to a wide plateau at several tens of nanometers [18, 20, 21]. It should be underlined that small-size clusters are absent in the wavelet spectra for all other variants of plasma treatment and, therefore, exposure duration has decisive importance [20]. This result should be taken into account in the further analyses.

3. **Analysis of Strong Fluctuations of Z Coordinate of the Investigated Surface**

Detection and separation of characteristic strong fluctuations of Z coordinate of the investigated surface were performed using self-organizing maps (SOM) by Kohonen [11-13] that ensure search of natural clusters without a priori conditions. Growing hierarchical self-organizing maps (GHSOM) were developed to detect complex data structures with hierarchical bonds [14]. High classification accuracy of this analysis method represents a wide range of options in studying morphology of thin films.

The algorithm principle is presented on FIG. 2, where GHSOM map is shown left. At the demonstrated analysis level GHSOM includes 4 classification boxes (2, 3, 4 and 5) of the separated clusters with maximum, minimum and interim film heights. The fourth box corresponds to the maximum heights area and the third one to the minimum heights area. The right part of this figure shows maps (map 2, map 3, map 4 and map 5) of cluster height distributions on the film obtained by means of unfolding relevant GHSOM data regarding X and Y coordinates of clusters.

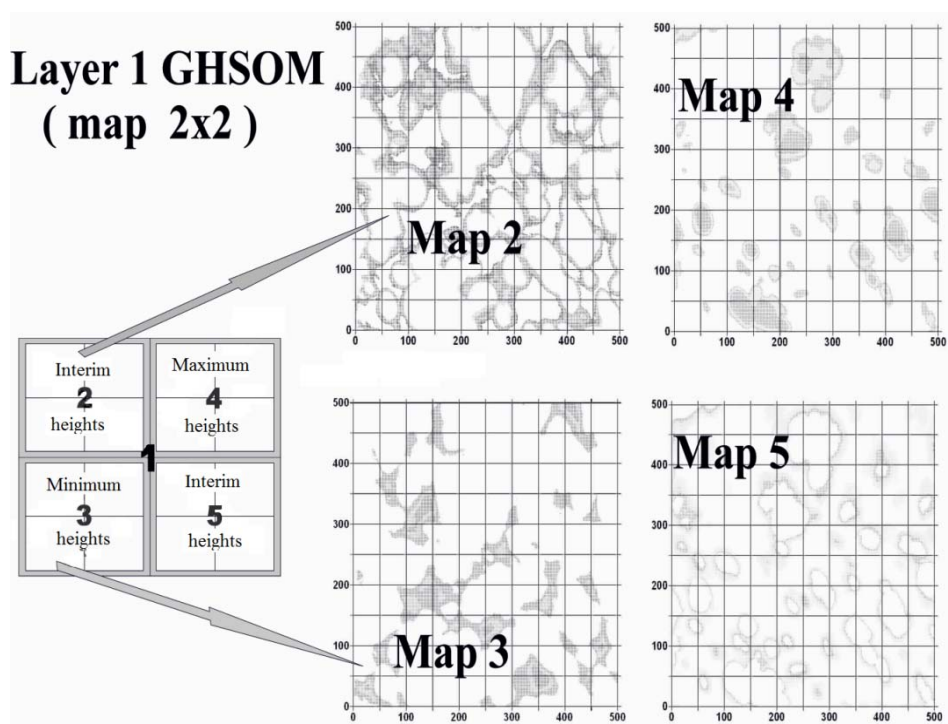

FIG. 2. Strong Z Fluctuations Analysis Scheme Using GHSOM

The obtained results are represented [18, 20, 21] in Table 1. It is obvious that morphological features of ZnO films have produced a clear parabolic dependence from duration of plasma treatment.

This parabola has the clear minimum 13.7 nm for the film treated by plasma within 10 min. The result enables efficient and quick to control the stages of process [19] aimed at achieving the desired film characteristics. In other words one can say that the scanning probe microscope and effective spectral and clustering data analysis methods can be very useful not only for analyses but also at designing of thin oxide film with optimal properties.

Table 1. Maximum Cluster Heights

| ZnO: duration of plasma treatment, min | Max Z, nm |
|---|---|
| 00 | 39.5 |
| 05 | 34.5 |
| 10 | 13.7 |
| 15 | 24.4 |

**4. Comparison of Analytical Results with Direct X-Ray and Optical Measurements**

The measurement of X-ray and optical transmission spectra provides all possible options to perform a full structural and size effect analyses of quite a complex hexagonal structure of wurtzite type ZnO films. The results of such analyses of oxide films have shown [18, 20, 21] the strong difference of the ZnO film with 10 min plasma treatment from the rest samples. The X-ray pattern of this film presented on FIG. 3(c) has a highly intensive new phase peak with Miller index (103), which corresponds to small-size crystallites. In a similar way the set of transmission spectra for all films on FIG. 4 shows [18, 20, 21] that specifically 10 min plasma treatment provides for the maximum transparency at the wave length ~ 500 nm and minimum transparency at the wave length ~ 1000 nm. Generally the results of X-ray investigations (FIG. 3) of all films show lines of the polycrystalline ZnO phase with well-shaped crystallites. Moreover two reflection lines from crystal planes with Miller indices (100) and (101) are found in all films evidencing stability of certain structural parameters of the zinc oxide.

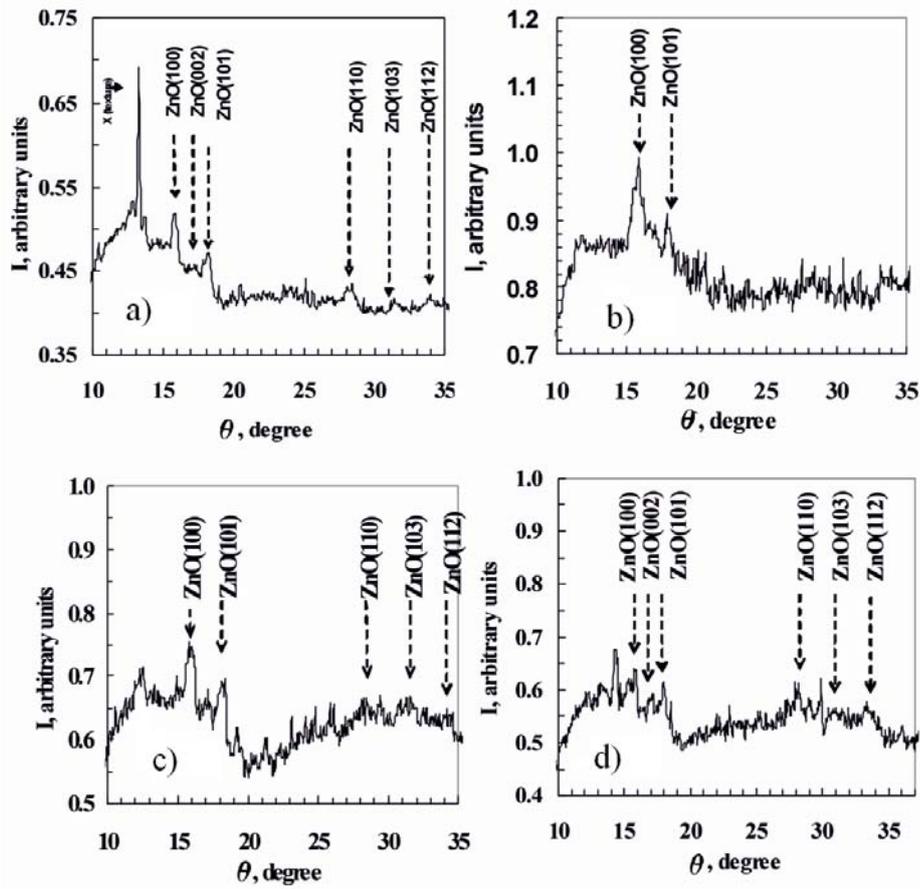

FIG. 3. X-Ray Patterns of ZnO Films Obtained by Sol-Gel Technique (a) and Modified by Hydrogen Plasma Treatment: 5 min (b), 10 min (c), 15 min (d)

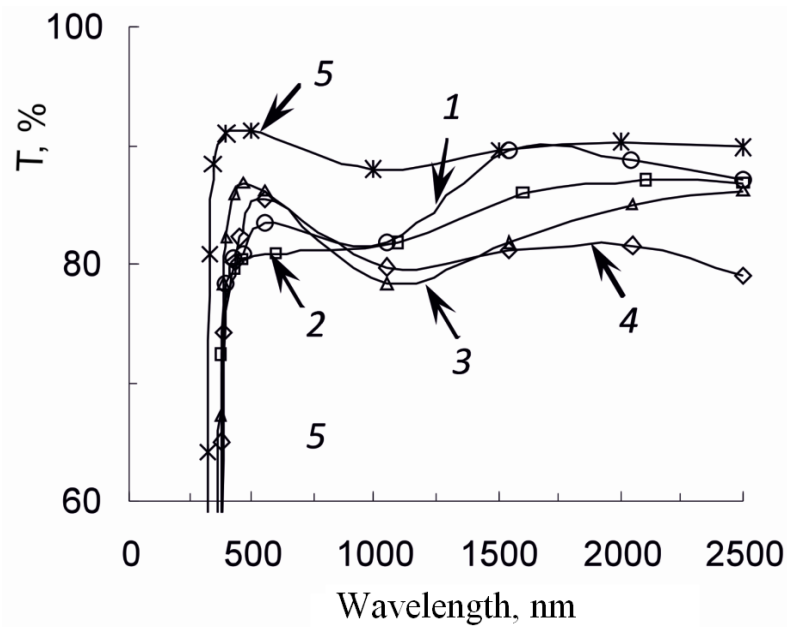

FIG.4. Optical Transmission Spectra of glass substrate (5) and thin ZnO Films Modified by Treatment of Hydrogen Plasma: after synthesis (1), after treatment - 5 min (2), 10 min (3), 15 min (4).

5. Discussion

Computational estimation of size-property relations in plasma-modified zinc oxide thin films can be characterized as the attempt to obtain qualitatively new level of the complex problem solution. Powerful spectral and clustering analyses of size and morphology film effects come to conclusion that plasma exposure duration has decisive importance. Indeed, all variants of analyses performed in this work have clear shown the strong difference of the zinc oxide thin film with 10 min plasma treatment from the rest films.

Only in this variant of plasma treatment scale pattern has the distinct narrow peak at approximately five nanometers, whereas the small-size clusters are absent in the wavelet spectra for all other films. The characteristic strong surface fluctuations of the investigated films build a distinct parabolic relation in which a minimum outlier also corresponds to the film with 10 min plasma treatment.

With direct X-ray measurements we have obtained that only for this film the X-ray spectrum has a highly intensive new phase peak with Miller index (103), which corresponds to small-size crystallites. Similarly, this film has been selected among all films by high transparency in the visible area of light.

The whole set of computational and experimental features of this work can be well explained from the point of view of plasma-induced phase transitions with critical point corresponds to the 10 min plasma treatment. It should be stressed, that one of the most striking features of phase transitions are the large fluctuations that become especially apparent near a critical point. Such approach gives understanding of our size-property relations as a function of duration of plasma treatment at the given quantity of power.

In the point of the phase transition, i.e. in the point of minimum fluctuations of Z coordinate, the continuous wavelet transform has recognized scale spectrum with the distinct narrow peak centered at approximately five nanometers. As mentioned before the control film sample and the films treated during 5 and 15 min do not contain small-size crystallites.

Self-organizing neuronet algorithm in the GHSOM variant has separated a distinct parabolic relation for the strong fluctuations of Z coordinates on the investigated film surfaces. The regular behavior of film heights open the way leading to the evaluation of optimal technological conditions for the film designing.

**Summary**

Size-property relations in plasma-modified ZnO thin films have been established as a function of plasma treatment duration. It has been shown that duration of plasma treatment is a key parameter of thin oxide modifications. The correlations between crystallite sizes and the morphological characteristics of films have been extracted on the basis of the wavelet and self-organizing map analyses of the scanning probe microscope (SPM) data matrices. The nanocluster structure of oxide films have been studied in detail with accuracy corresponding to the results of the SPM with a tenfold sharper cantilever tip. The strong plasma-induced changes in crystallite sizes have been interpreted as a size-structure phase transition. Direct measurements of X-ray diffraction and optical transmittance spectra have confirmed the computational results. The exploration of size-property relations has shown that SPM and effective spectral and clustering approaches can be very useful not only at analyses but also at optimal property designing of thin oxide films for promising nanophotonics devices.

---